# Ontology-based Mediation with Quality Criteria

Fahad Muhammad

Univ Lyon, Univ Lyon 2, UR ERIC, France

`f.muhammad@univ-lyon2.fr`

**Abstract.** This paper presents a semantic system named *OntMed* for an ontology-based data integration of heterogeneous data sources to achieve interoperability between heterogeneous data sources. Our system is based on the quality criteria (*consistency, completeness and conciseness*) for building the reliable analysis contexts to provide an accurate unified view of data to the end user. The generation of an *error-free global analysis context* with the semantic validation of initial mappings generates accuracy, and provides the means to access and exchange information in semantically sound manner. In addition, data integration in this way becomes more practical for dynamic situations and helps decision maker to work within more consistent and reliable *virtual data warehouse*. We also discuss our successful participation in the Ontology *Alignment for Query Answering (OA4QA)* track at OAEI 2015 campaign, where our system (DKP-AOM) has performed fair enough and became one of only matchers whose alignments *allowed answering all the queries of the evaluation*.

**Keywords:** Ontology based Mediation, Ontology, Quality Criteria

## 1 Introduction

*Information integration* has long history since human started using and collecting information. But, it has been focus of IT research since 1970s. It deals with providing a unified and transparent access to a collection of heterogeneous data sources. In information integration, the formulization of a global schema is a difficult task that manages multiple, autonomous and heterogeneous data sources. According to the study by Bernstein and Haas [1], large enterprises utilize at least the 40% of their budgets on the information integration, and a forecast about the market for the worldwide data integration and access software estimated to grow from $2.5 billion to $3.8 billion in 2007 to 2012 [2]. Although, there is a huge research done on this topic, but, still it is a hot issue in IT research. The two major research challenges are Intra-organization and Inter-organization information integration. *Intra-organizational data integration* is vital when different components of an organization adopt different systems to maintain. The need of Inter-organizational data integration is required in companies' merger, stock exchanges, etc. Research on this topic evolves with time from centralized system with single and multiple data stores to managing federated data sources. Then, decentralized systems were designed where distributed integration is done by the application with or without central global schemas. Therefore, in the research literature, there are various approaches for the information integration, such as Data exchange, Mediator-based, P2P data integration and exchange, Data Warehousing, etc., [3]. Data exchange that aims at materialization of the global view by providing facilities of query answering without accessing the local data sources. There can also be P2P data integration and exchange between several peers. The approach is designed such that it allows queries over one peer, such that each of the peers is equipped with local and external sources. This paper is focused on *mediator based data integration* or *virtual warehouse based on quality criteria*. *Virtual* means that the data remains at the local data sources, and a mediation layer provides *transparency* in answering querying by managing underlying local heterogeneous data sources.

The rest of the paper is organized as follows. Section 2 presents related work. In section 3, we presented our *ontology-mediator based data integration* system *(OntMed)*. Section 4 presents the evaluation of our system with the help of Ontology Alignment for Query Answering (OA4QA) track of Ontology Alignment Evaluation Initiative (OAEI). Finally, section 5 concludes our paper and shows future directions.

## 2 Related Work

This section discusses the existing approaches of mediation for querying heterogeneous ontologies. There exist three approaches for querying heterogeneous data sources in research literature. According to the survey on ontology based information integration by Wache et al. [4], existing approaches can be classified into three broad categories. These different categories for ontology based data integration are as represented in Figure 1 and elaborated below.

**Single Ontology Approach.** In a single ontology approach, all source schemas are directly related to a shared global ontology. The shared global ontology provides a uniform interface and facilitates querying to the end-user. SIMS is a system that exploits single ontology approach for querying heterogeneous ontologies [5].

**Multiple ontology Approach.** In multiple ontology approach, each local data source is represented by its own local ontology. Local ontologies are mapped to each other on basis of their similarities, i.e., their inter-ontology mappings are defined. The user query is rewritten for each local source and finally results are merged as per the inter-ontology mappings. Observer is a system that exploits multiple ontology approach for querying heterogeneous ontologies [6].

**Hybrid ontology Approach.** Hybrid ontology approach takes the advantages of both the above approaches [7]. For each local data source, there is a representative local ontology. All the local ontologies are mapped to a global shared ontology that provides a unified view. The advantage of this approach is that new data sources can be easily added with no need for modifying existing mappings.

On the basis of these approaches, mapping between the local and global ontologies are specified. When the source schemas are defined with respect to the global schema, the approach is term Local-As-View (LAV). But when the global schema is defined with respect to the local schema, the approach is known as Global-As-View (GAV). A mixed approach can also be defined known as GLAV.

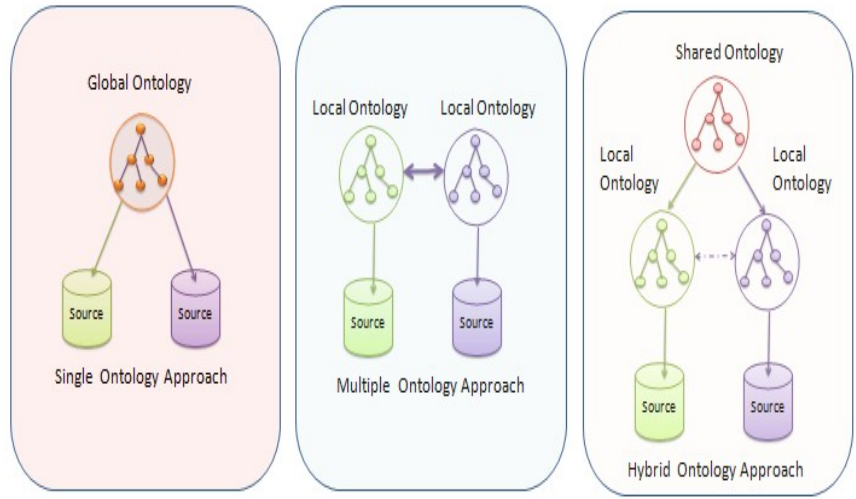

Figure 1. Different approaches for Ontology based Data Integration

Data integration is crucial since the analysis context is built using data from different heterogeneous data sources and providing an end user with the unified view of underlying data [8]. For the integration of heterogeneous data and on-line analytical processing, data warehousing has long been adopted that aims at building a centralized database which contains all (or selected) data coming from different data sources modeled in the multidimensional way [9]. Two well-known techniques for building warehouses are provided by *Inmon [10]* and *Kimball [11]*. In addition, on-line analytical processing (OLAP) systems may store terabytes of data and support queries from thousands of end-users. Since in the data warehouse, heterogeneous data is warehoused that facilitates the decision processes, this approach is characterized by its performance in terms of a query response time. But, when the data changes over time, decisional tools necessitate several updates for the sound decision making which need much additional cost for the refreshment of data. Data warehouse by mediation is thus increasingly seen as a solution to this problem that aims at building a virtual data warehouse to process analysis context on-the-fly.

Now-a-days, use of ontologies for describing semantics of data in data warehouse design has revolutionized the performance in terms of accuracy, and provides means to exchange information in semantically sound manners in heterogeneous environments. Although, there is already a lot of research in this area, there are still many issues that need to be resolved especially concerning the quality of results. Based on the previous research [12], we analyzed various situations when local ontologies are merged together that need consistency, completeness and conciseness checking. Therefore, we build a quality criteria based on *consistency, completeness and concise-* *ness* parameters for building the reliable analysis context in the data warehouse design based on a mediation approach. We further investigated our approach that exploits the semantic validation of initial mappings from the information present in the local ontologies that form global analysis context on-the-fly. Global ontology as an analysis context provides a conceptual unified view to the end-user about the underlying data. The user performs a query, over the data via the global ontology. As this query is expressed in a global schema terminology, it must be reformulated in terms of local data sources such that it can be executed. Otherwise, query results incomplete or null results. Once the results are retrieved from the global schema, they need to be merged together and presented to the user in terms of a global terminology. In this whole process, we believe that the generation of error free global analysis contexts with semantic validation of initial mappings would generate more accuracy, provide means to exchange information in semantically sound manners, data integration by this way becomes more practical for dynamic situations and helps decision maker to work within more consistent and reliable virtual data warehouse. There are many other problems under this topic, such as data extraction, cleaning, reconciliation, and optimization of query answers, but these are out of scope of this paper. Our work is limited to mapping identification between local data sources, global schema generated, and analyzing how to answer queries expressed on the global schema.

## 3 Ontology-based Data Integration

Traditional data warehouse systems based on *ETL* mechanisms *extract*, *transform*, and *load* data from several heterogeneous data sources into a single queryable data. This approach of data integration is tightly coupled because the data reside together in a single repository at query-time, and provides fast query response. As of modern research (almost from 2009 onward), the trend in data integration has favored loosening the coupling between the data. This requires mediation based uniform query-interface, where an end-user can pose a query, and the mediation layer transforms it into specialized queries over the original databases, and retrieves the results. One of an example of data warehouse based on mediation approach is presented by [13].

A data warehouse based on a mediator approach consists of three main elements; *data source schemas* as local ontologies, *a global ontology* and *correspondence rules* between local-global ontologies for query answering. Building a global ontology is essential as it allows the execution of decisional queries on underlying sources, query transformation from the global schema language to the data sources languages, and building the data cube on-the-fly from the obtained results from the different data sources. But, the mediation layer have to face number of problems, since local ontologies may represent the same knowledge in different ways producing various mismatches while the construction of a global schema. In addition, it is very hard to perform their integration manually beyond a certain degree of complexity, size, and number of ontologies, and need fully automatic methodologies for enabling interoperability in dynamic environment such as virtual data warehouse where analysis context is made on-the-fly. This dilemma requires the need of a reliable ontology merging ap-proach that should be capable enough to find semantic heterogeneity in source ontologies and resolve it automatically to produce accurate, consistent and coherent global merged ontology on-the-fly based on a number of local source ontologies. Since it is in the analysis and decision domain, so validation of initial mappings that form global schema is more than a challenging task for decision making and sound manipulation of data. Therefore, we took this initiative integrate a quality criteria inside mediation layer to build a reliable analysis context to achieve correctness of desired query results.

### 3.1 Representative Schema for Local Data Sources

We believe each of the data sources has some representative schema, i.e., local sources ontologies already exist. If not, use some tool to generate individual ontologies [14, 15].

### 3.2 Generation of Global Schema

The second step of ontology-based data integration is to formulate the global schema. Our approach is flavored by building a mediator based data integration that facilitates to physically reside data in original sources and frees the user from having to locate sources relevant to a query, interact with each one in isolation, and manually combine data from multiple sources. For our ontology-based data integration system, we used our previously designed ontology merging system DKP-AOM that deals with the problem of providing a unified view (in terms of global ontology) of local ontologies which represent distributed and heterogeneous data sources. Further information about our DKP-AOM system can be found in [16]. The user pose a query in terms of a global schema and the mediator adopts some mechanisms to execute the query to provide meaningful answers. For this, it needs to translate a query formulated on the mediated schema into one that refers directly to the local schema.

### 3.3 A Novel Framework for Data Integration with Reliable Mediation

We designed a novel framework for the data integration with reliable mediation named *OntMed* based on a hybrid approach; where local ontologies represent local data sources, our DKP ontology merging framework generates a global ontology automatically from the local ontologies, and finally a query engine executes query and retrieve result by query rewriting, obtains local answers, and finally constructs global result. Figure 2 depicts the data integration approach adapted by our system *OntMed*.

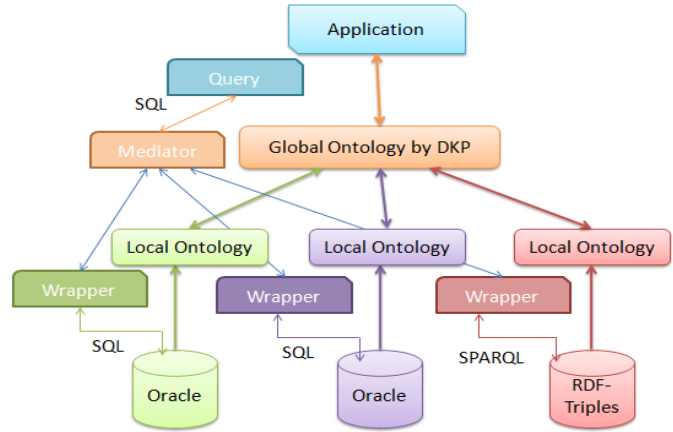

Figure 2. Data Integration approach adapted by OntMed

It is well argued in the research area that often the generated results of data integration suffer from inconsistency, incompleteness and redundancy. Therefore, we implemented the quality criteria based on the Consistency, Completeness and Redundancy parameters [17, 18] in a mediation layer to build an accurate analysis context so that execution of simple queries on the global ontology reveal correct results. The details about these criteria are as follows.

**Inconsistency in the global ontology.** Inconsistency in the global ontology means that there is some sort of contradictory knowledge inferred from the concepts, definitions and instances within the ontology. It creates ambiguity, contradiction in results and compromises precision. There are mainly three types of errors that can cause inconsistency and ambiguity in the ontology. These are Circulatory errors, Partition errors and Semantic inconsistency errors.

**Incompleteness in the global ontology.** Incompleteness occurs when the global ontology contains various types of domain knowledge in the form of concepts, properties and definitions, but, overlooks some of the important information about the domain. The incompleteness of the domain knowledge often creates ambiguity, and lacks reasoning and inference mechanisms. The incompleteness errors are due to incomplete concept specification and partitions errors due to disjointness and exhaustive knowledge omission between concepts.

**Redundancy in the global ontology.** Another important aspect is to make global ontology concise, so that it stores only necessary and sufficient knowledge about the concepts. Redundancy errors not only compromise conciseness and usability, but also create problems for the maintenance and manageability of ontologies. Generally, redundancy occurs when particular information is inferred more than once from the relations, classes and instances found in the global ontology.

**Design Anomalies in the global ontology.** Besides taxonomic errors, Baumeister and Seipel identified some design anomalies that prohibit simplicity and maintainability of taxonomic structures within the ontology [19]. These do not cause inaccurate reasoning about the concepts, but point to problematic and badly designed areas in ontology that cause issues of maintainability. Identification and removal of these anomalies should be necessary for improving the usability, and providing better maintainability of online ontologies over the semantic web.

Figure 3 depicts our quality criteria for achieving reliable mediation and also summaries various errors. The important thing which should be considered is that global ontology should be free from these errors as it serves an analysis context on which user queries are performed and final result is aggregated.

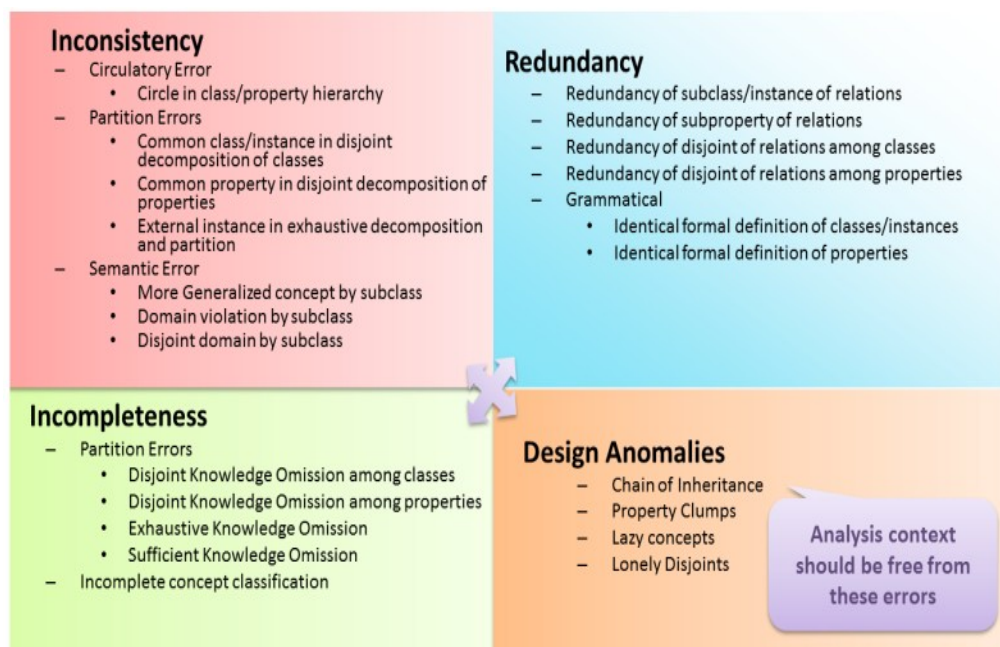

Figure 3.Quality criteria for Reliable Mediation

Once the global ontology is generated by DKP ontology merging system, this serves as analysis context for our data integration system. We have participated in the conference and OA4QA track of OAEI 2015 and results of our system are competitive see [20]. Figure 4 shows the execution of different steps of our system for the data integration diagrammatically. We have tested our approach of data integration on a case study related to ShoppingMall data sources which are modeled in two instances of Oracle. These data sources are heterogeneous and separately developed. In addition, we blend our case study with the RDF triple data source, to add more heterogeneity environment in the case study. However, we have incorporated very limited information about the RDF data source in this case study, and mainly it circulates over Oracle data sources. In near future, we will present this case study.

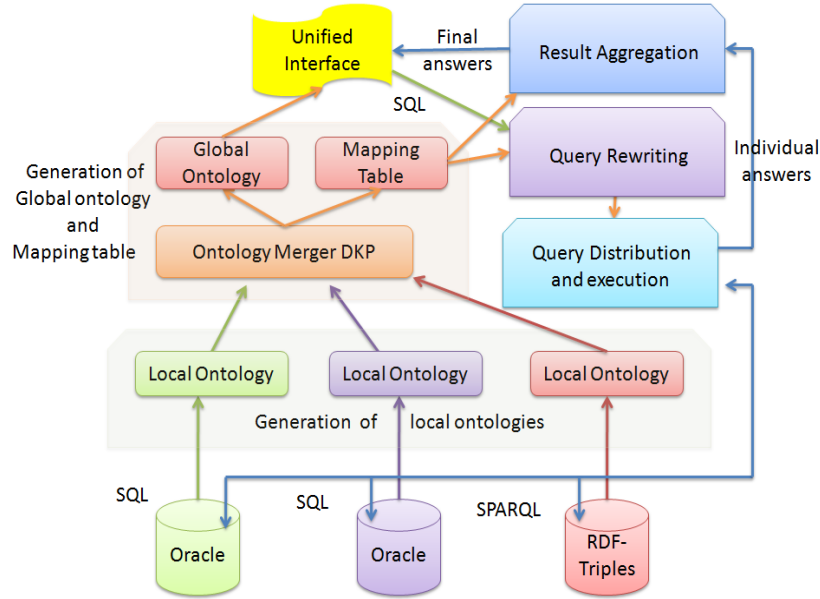

Figure 4. Execution of different steps by Ont-Med

## 4    Validation of OntMed Methodology at OAEI-2015

We have implemented OntMed (ontology-based data integration) module as part of a Ontology Merging System (DKP-AOM). For the evaluation of our system, we choose OBDA scenario (see Figure 5) part of Ontology Alignment for Query Answering (OA4QA) track of Ontology Alignment Evaluation Initiative (OAEI) 2005 [21]. Although it is very old, but serves its purpose to validate the methodology. Three principles are taken into account to minimize the number of potentially unintended consequences [22], namely: (i) *consistency* principle, the alignment should not lead to unsatisfiable classes in the integrated ontology; (ii) *locality* principle, the correspondences should link entities that have similar neighborhoods; (iii) *conservativity* principle, the alignments should not introduce alterations in the classification of the input ontologies.

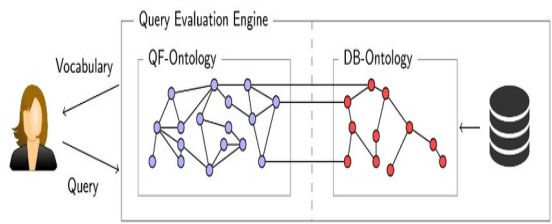

Figure 5. Ontology-based data access scenario

Consider an OBDA scenario where one ontology provides the vocabulary to formulate the queries (QF-Ontology) and the other is linked to the data and it is not visible to the users (DB-Ontology).The integration via ontology alignment is required between these two ontologies since only the vocabulary of the DB-Ontology is connected to the data. Given a query and an ontology pair, a model (or reference) answer set will be computed using the correspondent reference alignment for the ontology pair. Precision and Recall will be calculated with respect to these model answer sets [23]. For this track, ontology alignment evaluation with respect to a set of reference alignments are taken into account investigating this aspect of data integration. Initiative campaign calculated Precision and recall with respect to the generated alignments to answer a set of queries in a ontology-based data access scenario where several ontologies exist. In addition, the track is intended as a possibility to study the practical effects of logical violations affecting the alignments, and to compare the different repair strategies adopted by the ontology matching systems. In order to facilitate the understanding of the dataset and the queries, the conference data set is used, extended with synthetic ABoxes. The result of participated systems are taken into account by SEALS platform.

| Matcher | Answered queries | ra1 | | | rar1 | | |
|---------|------------------|-------|------|------|-------|------|------|
| | | Prec. | F-m. | Rec. | Prec. | F-m. | Rec. |
| AML | 18/18 | 0.78 | 0.76 | 0.75 | 0.76 | 0.75 | 0.75 |
| LogMap | 18/18 | 0.75 | 0.75 | 0.75 | 0.73 | 0.73 | 0.73 |
| XMAP | 18/18 | 0.78 | 0.72 | 0.68 | 0.72 | 0.70 | 0.67 |
| LogMapC | 18/18 | 0.72 | 0.71 | 0.69 | 0.72 | 0.71 | 0.70 |
| COMMAND | 14/18 | 0.72 | 0.66 | 0.61 | 0.69 | 0.62 | 0.56 |
| DKPAOM | 18/18 | 0.67 | 0.64 | 0.62 | 0.67 | 0.66 | 0.65 |
| Mamba | 14/18 | 0.71 | 0.61 | 0.53 | 0.71 | 0.61 | 0.54 |
| CroMatcher | 12/18 | 0.70 | 0.57 | 0.48 | 0.61 | 0.49 | 0.4 |
| LogMapLt | 11/18 | 0.70 | 0.52 | 0.42 | 0.58 | 0.43 | 0.35 |
| GMap | 9/18 | 0.65 | 0.49 | 0.39 | 0.61 | 0.43 | 0.33 |
| Lily | 11/18 | 0.64 | 0.47 | 0.37 | 0.64 | 0.48 | 0.39 |
| JarvisOM | 17/18 | 0.43 | 0.43 | 0.43 | 0.43 | 0.41 | 0.39 |
| ServOMBI | 6/18 | 0.67 | 0.33 | 0.22 | 0.67 | 0.33 | 0.22 |
| RSDLWB | 6/18 | 0.39 | 0.25 | 0.18 | 0.39 | 0.19 | 0.13 |

Table 1. Results published by OA4QA track ***taken from original paper [21]

Results for the whole set of queries shows the average precision, recall and f-measure (see Table 1). Matchers are evaluated on 18 queries in total, for which the sum of expected answers is 1724. Some queries have only 1 answer while other have as many as 196. Our system DKPAOM is one of other systems (AML, LogMap, LogMap-C and XMap) which were the only matchers whose alignments allowed answering all the queries of the evaluation. This is our first successful participation in OAEI 2015.

## 5    Conclusion

Data integration is crucial since the analysis context is built using data from different heterogeneous sources. With the grown usage of data warehouse, the question about the usage of a mediation approach for building analysis context on-the-fly has become even more important in today's dynamic world. Although, there has been already a lot of research in this area, there are still many issues that need to be resolved especially concerning with the quality of results. There are various points about the ontology merging and conceptual schema merging. In general, an ontology is a broad concept than conceptual schema and a conceptual schema is regarded a sub-concept of ontology. Therefore, merging conceptual schema requires less effort than ontology merging. Integration of databases has been the focus of years of research. There are many approaches in the research literature for the data integration with/without the support of ontologies. Ontologies conceptualize concepts with their generalization and formulate properties and axioms to equip the semantics. The relational model on the other hand provides limited semantic description about the data. For example, there is no or very limited generalization support, no axiomatic definitions, etc. An absence of the rich semantics also poses a difficulty in recognition of concepts during their mappings and hence their merging raises the level of difficulty.

This paper contributes an ontology-based data integration framework. Integration of data sources also depends on the representation and quality of its representative schema or ontology. The more representation model is good, the more integration is easily done with quality. The presented approach based on the quality criteria (consistency, completeness and conciseness) for building reliable analysis contexts in data warehouse design suits well and provides accurate unified view of data to the end user. In addition, our approach builds the virtual data integration environment with the least human intervention. Our automatic ontology merger generates the global ontology from the local ontologies. Our approach exploits semantic validation of initial mappings from the knowledge that is present inside the local ontologies that form global analysis context on-the-fly. The generation of an error-free global analysis contexts with the semantic validation of initial mappings generates more accuracy, and provides means to exchange information in semantically sound manners. In addition, data integration by this way becomes more practical for dynamic situations and helps decision maker to work within more consistent and reliable virtual data warehouse.